# Temperature and Gas-Environment Dependent Electron and Phonon Transport in Suspended Carbon Nanotubes Up to Electrical Breakdown

David Mann,[1] Eric Pop,[1,2] Jien Cao,[1] Qian Wang,[1] Kenneth Goodson[2] and Hongjie Dai[1,*]

[1] *Department of Chemistry and Laboratory for Advanced Materials, Stanford University, Stanford, CA 94305, USA*

[2] *Department of Mechanical Engineering and Thermal Sciences, Stanford University, Stanford, CA 94305, USA*

High bias electrical transport characteristics of freely suspended metallic single-walled carbon nanotubes (SWNTs) are investigated at 250-400K in vacuum and various gases. Self-heating is exploited to examine the temperature dependence of phonon transport and optical phonon decay in SWNTs. The acoustic phonon thermal conductivity of a SWNT follows ~ $1/T$ at high temperatures. Non-equilibrium optical phonon effects in suspended nanotubes decrease as the ambient temperature increases. Gas molecules assist the relaxation of hot optical phonons along the tube length and enable enhanced current flow. In vacuum, high bias breakdown of suspended SWNTs can occur via melting caused by electrically emitted hot optical phonons at a low acoustic phonon temperature.

* E-mail: hdai@stanford.edu.  DM and EP contributed equally.

The effect of the environment on the current carrying ability of nanostructures is important for the performance optimization of next-generation electronic devices and interconnects. It has been shown recently[1] that freely suspended SWNTs display negative differential conductance (NDC) and sharply reduced current levels compared to nanotubes on substrate[2,3] due to substantial self-heating and scattering with non-equilibrium optical phonons (OPs) emitted under high-bias current flow.[1] This is an extreme case of high field electrical transport in one dimensional (1-D) materials which is dependent on the environment surrounding SWNTs. The study of electro-thermal transport is also generally important in 1D nanostructures due to the reduced dimensionality and its effect on thermal conduction and phonon decay.

In the current work, we investigate the effects of several environmental factors on electro-thermal transport in suspended SWNTs, including ambient temperature, gas pressure and species. A new electrically driven melting phenomenon involving hot OPs and relatively cold acoustic phonons (ACs) is also observed. Suspended ($L \sim$ 1-3 $\mu$m long, $d \sim$ 1.8-3 nm diameter) metallic SWNTs were obtained by CVD growth across pre-formed trenches and Pt electrodes (Fig. 1) as described previously.[4,5] The devices were characterized by scanning electron microscopy (SEM) to obtain nanotube length information. Electrical measurements were carried out in an environmental probe station between temperatures $T_0$ = 250-400K in vacuum ($\sim 10^{-6}$ torr) and various gases.

We have measured large numbers of individual suspended SWNTs at various $T_0$ and always observed their hallmark negative differential conductance (NDC) behavior at high bias (Fig. 2a).[1] Interestingly, the $I$-$V$ curves taken at different $T_0$ tend to converge in the high bias regime (Fig. 2a). To interpret the temperature-dependent $I$-$V$ characteristics, we adopt and extend a recently introduced electro-thermal model.[1] The resistance of a SWNT under self-heating can be written as

$$R(V,T) = R_c + \frac{h}{4q^2} \frac{L + \lambda_{eff}(V,T)}{\lambda_{eff}(V,T)}, \qquad (1)$$

where $R_c$ is the contact resistance and $\lambda_{eff} = \left(1/\lambda_{ac} + 1/\lambda_{op,ems} + 1/\lambda_{op,abs}\right)^{-1}$ is the bias and temperature dependent electron mean free path (MFP), including scattering with AC phonons, and OP emission and absorption.[1] The current is computed as $I=V/R$. The lattice temperature distribution along the SWNT is obtained by solving the heat conduction equation with $T=T_0$ boundary conditions at both ends of the tube[6]

$$A\nabla(\kappa_{th}\nabla T) + p' - g(T - T_0) = 0, \qquad (2)$$

where $\kappa_{th}$ is the SWNT thermal conductivity, $p' = I^2(R-R_c)/L$ is the Joule heating per unit length, $A=\pi db$ is the cross-sectional area ($b \sim 0.34$ nm the tube wall thickness) and $g$ is the net heat loss (by radiation and conduction to ambient gas) per unit length. We solve Eqs. (1) and (2) iteratively until the temperature profile at each point along the SWNT converges within 0.1 K, at every bias. This profile corresponds to the AC phonons ($T_{ac}$), which are the main heat carriers in the temperature range considered.[7] However, since Joule heating at high bias is initially dissipated to OP modes with long lifetimes in suspended SWNTs, a non-equilibrium OP population develops.[1] We capture this non-equilibrium situation with an effective OP temperature

$$T_{op} = T_{ac} + \alpha(T_{ac} - T_0) = 0 \qquad (3)$$

where $\alpha > 0$. The captures the physical picture of OP relaxation by decay into AC modes ($T_{op}$ to $T_{ac}$) which are subsequently carried out of the tube through the contacts ($T_{ac}$ to $T_0$). OP decay by direct propagation out of the tube is not appreciable due to their small group velocity. Eqs. (1)-(3) form the basis of our electro-thermal transport model and are used to calculate $T_{ac}$ and $T_{op}$ along the tube, the respective AC and OP scattering MFPs, and consequently the R-V and I-V characteristics of the nanotube.

In vacuum where $g \sim 0$, Joule heating in the tube dissipates along its length to the contacts. This results in a nearly parabolic temperature profile along the SWNT. We find that for a $\kappa_{th}(T) \sim 1/T$ dependence (consistent with Umklapp phonon scattering at high temperatures),[1,8] our model enables good reproduction of the experimental I-V curves (Fig. 2a) and numerical extraction of the non-equilibrium coefficient $\alpha$ (Fig. 2b). We find the non-equilibrium coefficient scales as $\alpha \sim 2.3(300/T_0)^m$ with $m \sim 1.5$ (Fig. 2b). This empirically observed relationship suggests increased non-equilibrium OP effects at lower $T_0$, which are responsible for the convergence of high-bias I-V curves recorded at various ambient temperatures (Fig. 2a).

We give a theoretical account for $\alpha(T)$ as follows. The average AC temperature can be written as $T_{ac} = T_0 + P\mathcal{R}_{th}$ and the average OP temperature as $T_{op} = T_{ac} + P\mathcal{R}_{op}$, where $\mathcal{R}_{op}$ and $\mathcal{R}_{th}$ are the *thermal* resistances for OP decay into AC phonons and for AC heat conduction along the tube, respectively. The Joule power $P$ is dissipated first to the OP modes which then decay into AC modes, a sensible assumption for suspended SWNTs under high bias, when transport is limited by high-energy ($\hbar\omega_{op} \sim 0.18$ eV) OP scattering.[1] Consequently the OP non-equilibrium coefficient $\alpha$ can be written as

$$\alpha = \frac{\mathcal{R}_{op}}{\mathcal{R}_{th}} \sim \frac{\tau_{op}/C_{op}}{L/(12A\kappa_{th})} \sim \left(\frac{12A}{L}\right)\left(\frac{\tau_{op}\kappa_{th}}{C_{op}}\right). \qquad (4)$$

The OP lifetime $\tau_{op} \sim 1/T$ scales approximately as the phonon occupation[8] and is consistent with previous observations of OP Raman linewidths $\propto T$.[9] The thermal conductivity $\kappa_{th} \sim T$ at low $T$ (in 1-D) and $1/T$ at high $T$,[8] and the OP heat capacity $C_{op} \sim T$ at low $T$ and approaches a constant at high $T$.[8] These lead to an expected temperature dependence of $\alpha \sim 1/T$ at low $T$ and $1/T^2$ at high $T$ (see trend lines in Fig. 2b). Our empirically observed $\alpha \sim 1/T_0^{1.5}$ falls within the theoretically expected trends (Fig. 2b). The lack of quantitative agreement is attributed to a non-constant temperature profile along our suspended tube. These results represent the first time that ambient temperature dependent non-equilibrium OP effects on high-bias transport in SWNTs have been investigated. The reduced non-equilibrium effect at higher $T_0$ is owed to stimulated OP to AC decay at higher temperatures and the $\kappa_{th} \sim 1/T$ AC thermal conductivity (Eq. 4).

We have also measured suspended SWNTs in various gases and pressures and observed an increase of the high-bias current when compared to vacuum (Fig. 3). This allows us to investigate how surrounding gas molecules affect conduction heat loss ($g$ in Eq. 2) and OP non-equilibrium ($\alpha$) in suspended SWNTs. We estimate an upper limit of $g \sim (nv/4)(3k_B/2)(\pi d) = 0.000428$ WK$^{-1}$m$^{-1}$ in 1 atm of N$_2$, where $n$ is the density and $v$ is the average velocity (dependent on gas type via molecular mass) of gas molecules.[10] This is much smaller than $g \sim 0.1$ WK$^{-1}$m$^{-1}$ estimated for tubes on substrates[6] and cannot explain the enhanced current observed at high bias assuming no gas effect on $\alpha$ and on the non-equilibrium OP lifetimes. The combined SWNT heat loss to the gas ambient and to radiation is estimated to be < 2 % of the power dissipation at all relevant temperatures. On the other hand, when allowing a varying $\alpha$ from its value in vacuum, our calculations successfully reproduce the $I$-$V$ curves of suspended SWNTs in various N$_2$ pressures (Fig. 3a). A systematic decrease in $\alpha$ is found as the gas pressure increases, indicating reduced non-equilibrium OPs. This suggests that gas molecules stimulate the relaxation of OPs in suspended SWNTs, most likely via coupling mechanisms involving vibrational or other degrees of freedom of molecular motion. The reduced $\alpha$ and $T_{op}$ decreases scattering with non-equilibrium OPs and enables enhanced high-bias current for nanotubes suspended in a gas ambient. We note that the *low bias* resistance of the suspended tubes is not affected by the presence of gases studied here, indicating negligible gas doping effect.[11]

We have previously suggested that for SWNTs lying on solid substrates, the OP lifetime is sharply reduced by the presence of additional phonon decay channels through the substrate/tube interface.[1] Such short lifetimes imply $\alpha \sim 0$ ($T_{op} \sim T_{ac}$) and no significant non-equilibrium OP effects for tubes on substrate, whose high bias $I$-$V$ characteristics are modeled with $g \sim 0.1$ WK$^{-1}$m$^{-1}$ due to heat sinking by the substrate.[6] Suspended SWNTs in vacuum represent the other extreme with large $\alpha \sim 2.4$ and $g \sim 0$. When in a gas ambient, the suspended tube exhibits a small $g$ but a non-equilibrium $\alpha$ in between that in vacuum and on a solid substrate. We have also measured the $I$-$V$ characteristics of suspended SWNTs in air, He and Ar (Fig. 3b). We find that monatomic gases have less influence on the high bias current, while diatomic gases such as $N_2$ and air ($I$-$V$ curve in air is indistinguishable from $N_2$) have a stronger influence. This suggests the gas effect on high-bias conduction is a result of gas molecules physisorbed on the surface of the SWNT. These molecules act as weak "substrates" (analogous to, but much weaker than $SiO_2$ in the case of tubes on such solid substrates), introducing additional phonon decay channels and leading to reduced OP lifetimes (decreased $\alpha$) along the tube. Diatomic gases are more effective than monatomic gases very likely due to the coupling of nanotube OPs with the vibrational modes of the gas.[12, 13] Notably, coupling between the vibrational modes of physisorbed molecules and surface OPs of solids have been previously documented,[12, 13] but not for SWNTs. We note that inert gases have been shown to affect the low bias electrical properties of macroscopic films of SWNT bundles due to indentation effects,[16] which differ from the enhancement of high bias electro-thermal transport properties of individual suspended SWNTs presented here.

Lastly, we explored the electrical breakdown of suspended SWNTs in vacuum. We find that suspended SWNTs in the $L \sim 1.8$-3.1 µm range consistently break at bias voltages between 2.4-3 V (Fig. 4a), and that the break appears to be occurring near the center of the tube (Fig. 4c). By comparing the $I$-$V$ curves (Fig 4a) using our self-heating model, we obtain estimates of the peak $T_{op}$ and $T_{ac}$ in the center of the SWNT and find they reach ~2200 K and ~800 K respectively (Fig. 4b). Notably, $T_{op} \sim 2200$ K is in the vicinity of the expected melting temperature of SWNTs,[14, 15] suggesting that electrical breakdown of a suspended SWNT in vacuum is due to non-equilibrium OP-mediated melting. Hot OPs emitted by electron scattering cannot rapidly decay and the short-wavelength energetic OPs are strong and long lasting to cause C-C bond breakage.

In summary, we have revealed the temperature dependence of non-equilibrium OP effects in suspended SWNTs in vacuum. Gas molecules surrounding a suspended SWNT are found to facilitate OP decay along the tube length, reduce the non-equilibrium OP effect and enhance the current carrying ability of suspended tubes. Interestingly, a suspended SWNT in vacuum can break down under high bias due to non-equilibrium OP-mediated melting at a relatively low acoustic phonon temperature.

We thank Prof. W. Harrison for insights. Project supported by MARCO MSD.

**Figure Captions:**

**Figure 1.** Suspended nanotube devices. (a) SEM image of a device showing a trench and source (S) and drain (D) electrode structures. (b) Close-up SEM of a SWNT suspended over the trench. (c) Schematic drawing of the device. Metallic and quasi-metallic nanotubes in on-states under a high negative gate voltage are used for all measurements in this work.

**Figure 2.** Self-heating of a suspended nanotube in vacuum. (a) $I$-$V$ characteristics of a SWNT with $L{\sim}2.6$ μm, $d{\sim}1.7$ nm and $R_c{\sim}8$ kΩ at four ambient $T_0$: experimental data (symbols) and model calculations (lines) based on SWNT thermal conductivity $\kappa_{th}(T) \sim 3400(300/T)$ Wm$^{-1}$K$^{-1}$. A comparison of our model with the data at high bias ($> 0.3$ V) allows the extraction of the OP non-equilibrium coefficient $\alpha$ at each $T_0$. (b) Numerically extracted OP non-equilibrium coefficient $\alpha$ (symbols) from (a) and trend lines as a function of contact/ambient temperature $T_0$. Decreasing $\alpha$ indicates reduced non-equilibrium OP effects at higher ambient temperatures.

**Figure 3.** Self-heating and electrical characteristics of a suspended SWNT in vacuum and various ambient gases. (a) Experimental $I$-$V$ data (symbols) and model calculations (lines) of a nanotube device ($L{\sim}2.1$ μm, $d \sim 2.4$ nm and $R_c \sim 4$ kΩ) in several ambient N$_2$ pressures. The gas pressure-dependent data can be reproduced by varying $\alpha$ roughly linearly with pressure. (b) $I$-$V$ data of the same tube as in (a) in various ambient gases all at 1 atm. pressure. We note that He has 5-10 times higher thermal conductivity (and $g$ value) than other gases, yet the high bias current does not correlate with this heat loss. The presence of He has the least influence on hot phonon decay with the largest remnant $\alpha$, whereas heavier and diatomic gases show a stronger effect on OP lifetimes ($\alpha_{He} \sim 2 > \alpha_{Ar} > \alpha_{N2} \sim 1.3$).

**Figure 4.** Electrical breakdown of suspended SWNTs. (a) $I$-$V$ data of two SWNTs up to breakdown in vacuum. The top curve is for a SWNT device with $L \sim 1.8$ μm, $d \sim 2.6$ nm, $R_c \sim 35$ kΩ and the bottom curve is for a SWNT device with $L \sim 3.1$ μm, $d \sim 1.8$ nm, $R_c \sim 75$ kΩ. Symbols are experimental data and lines are the calculations. Note that shorter suspended tubes break down at higher biases than longer ones. (b) Calculated peak optical phonon and acoustic phonon temperatures for the $L \sim 1.8$ μm SWNT. The $L \sim 3.1$ μm tube breaks at a lower voltage upon reaching the same peak $T_{op} \sim 2200$K (plot not shown). (c) SEM image of a device after electrical breakdown. Arrows point to the remaining end-segments of the broken tube that have sprung back to the electrode terraces after breakdown. The sample was coated with 1.5nm Ti/2.5nm Au to facilitate SEM imaging.

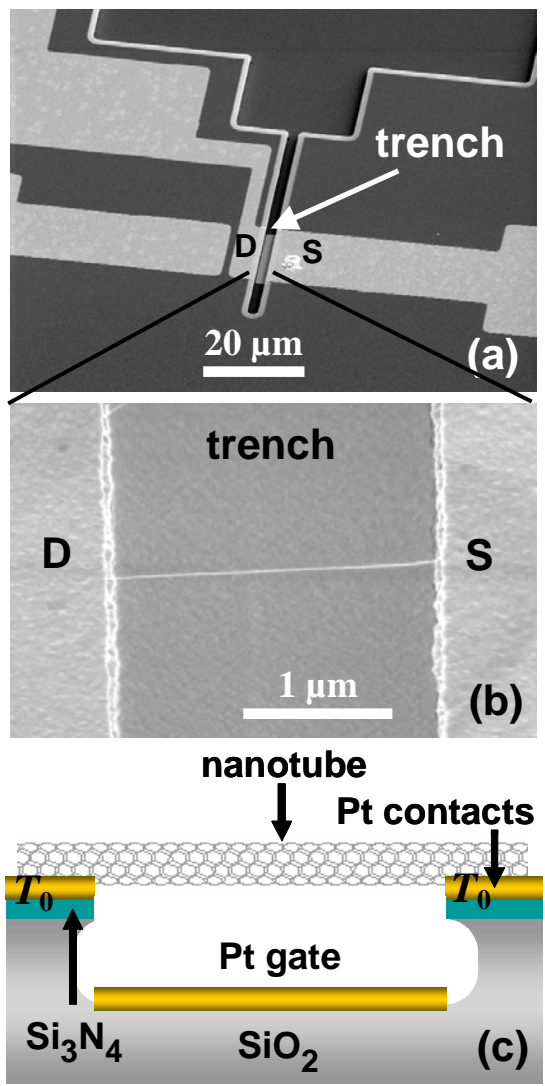

**Figure 1.**

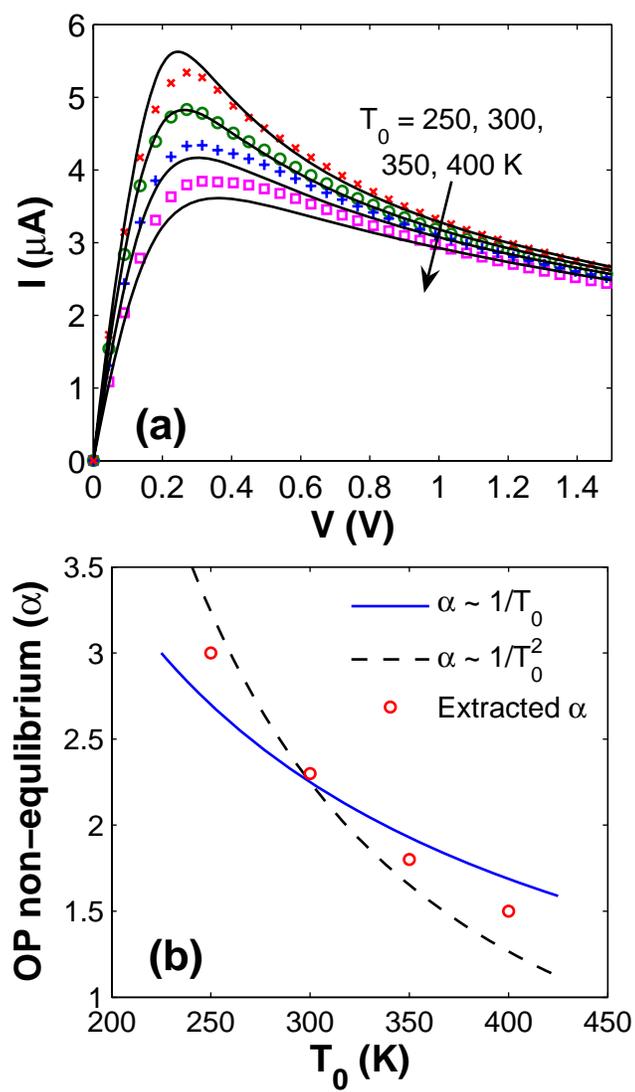

**Figure 2.**

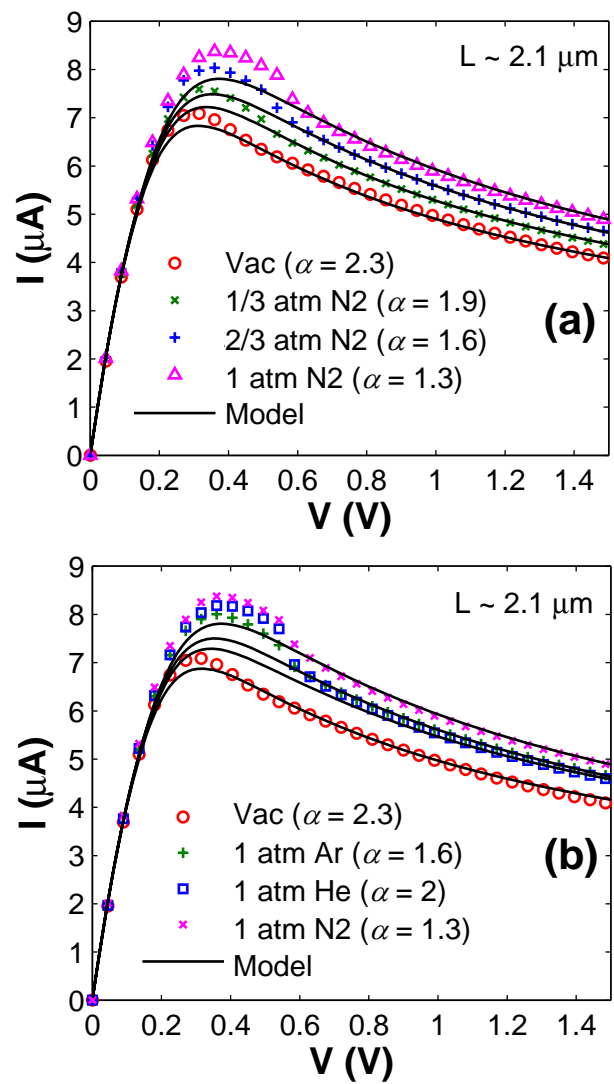

**Figure 3.**

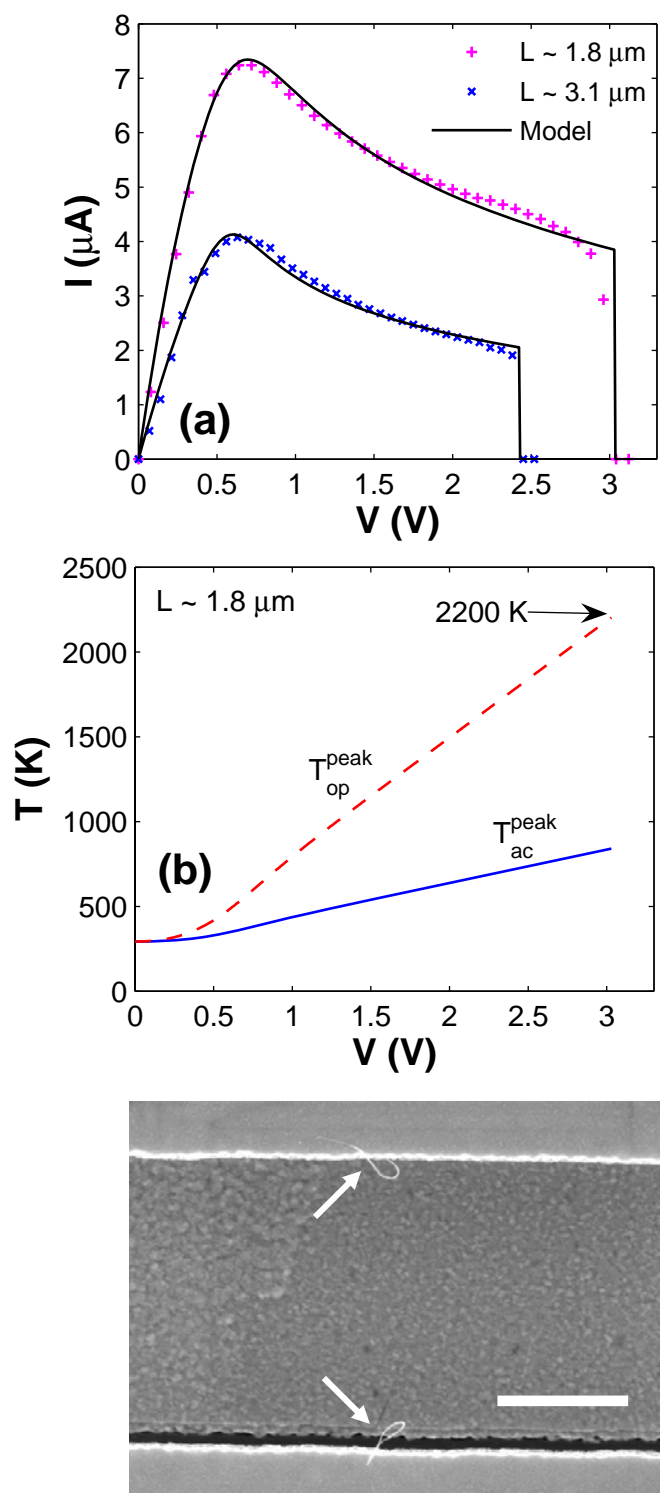

**Figure 4.**